\documentclass[twocolumn,showpacs,preprintnumbers,amsmath,amssymb,aps,prd,nofootinbib,superscriptaddress,eqsecnum]{revtex4}
\usepackage{makeidx}
\usepackage[dvipdfmx]{graphicx}
\usepackage[all]{xy}
\usepackage{feynmp}
\usepackage{amsmath,amssymb,bm}

\begin{document}

\title{Enhancement of Chiral Symmetry Breaking from the Pion condensation at finite isospin chemical potential in a holographic QCD model}

\author{Hiroki Nishihara~\footnote{e-mail: h248ra@hken.phys.nagoya-u.ac.jp}}
\author{Masayasu Harada\footnote{e-mail: harada@hken.phys.nagoya-u.ac.jp}}
\affiliation{
Department of Physics, Nagoya University, Nagoya 464-8602, Japan
}

\date{\today}

\def\theequation{\thesection.\arabic{equation}}
\makeatother

\begin{abstract}
We study the pion condensation at finite isospin chemical potential using a holographic QCD model.
By solving the equations of motion for the pion fields together with those for the iso-singlet scalar and iso-triplet vector meson fields, we show that the phase transition from the normal phase to the pion condensation phase is second order with the mean field exponent, and that the critical value of the isospin chemical potential $\mu_I$ is equal to the pion mass,
consistently with the result 
obtained by the chiral effective Lagrangian at $O(p^2)$.
For higher chemical potential, we find a deviation, 
which can be understood as a higher order effect in the chiral effective Lagrangian.
We investigate the $\mu_I$-dependence of the chiral condensate defined by
$\tilde{\sigma} \equiv \sqrt{ \langle \sigma \rangle^2 + \langle \pi^a \rangle^2 }$.
We find that $\tilde{\sigma}$ is almost constant in the small $\mu_I$ region, while it grows
with $\mu_I$ in the large $\mu_I$ region.
This implies that the strength of the chiral symmetry breaking is not changed
for small $\mu_I$:
The isospin chemical potential plays a role to rotate
the ``vacuum angle'' of the chiral circle 
$\tan^{-1} \sqrt{  \langle \pi^a \rangle^2 / \langle \sigma \rangle^2 } $
with keeping the ``radius'' $\tilde{\sigma}$ unchanged for small $\mu_I$.
For large $\mu_I$ region, on the other hand,
the chiral symmetry breaking is enhanced by the existence of $\mu_I$.
\end{abstract}


\maketitle

\section{Introduction}
\label{sec:Introduction}

Quantum ChromoDynamics (QCD) at finite isospin chemical potential is an interesting subject to study.
It combined with the finite baryon number chemical potential will provide a clue to understand the symmetry energy which is important to describe the equation of state inside neutron 
stars~\cite{Lattimer:2006xb}. 
In addition, it may give some informations on the structure of the chiral symmetry 
breaking~\cite{Friman:2011zz}.

When we turn on the 
isospin chemical potential $\mu_I$ at zero baryon number density,
the pion condensation is expected to occur at a critical point.
Son and Stephanov \cite{SS} showed that,
using the chiral Lagrangian at $O(p^2)$,
the
phase transition to the pion condensation phase is of the 
second order and the critical value of $\mu_I$ is equal to the pion mass.
It was also shown~\cite{Son:2000by} that $\langle \bar q \gamma_5 q \rangle$ condenses in the high isospin density limit.
Then a conjecture of no phase transition from the pion condensation phase to $\langle \bar q \gamma_5 q \rangle$ condensation phase was made.
The structure in the mid $\mu_I$ region is highly non-perturbative issue, so that it is not easy to understand such a region.

A pure isospin matter with zero baryon density can be simulated by the lattice analysis.
References~\cite{Kogut:2002tm,Kogut:2002zg,Detmold:2012wc} shows that the phase transition is of the second order, and that the critical chemical potential is equal to the pion mass.
Due to the existence of the sign problem, it is difficult to apply the lattice analysis for studying the hadron property at the finite baryon number density.
In this sense,
analysis by models may give some clues to understand the phase structure and the relevant phenomenon 
in the mid $\mu_I$ region. 
Actually, many analyses were done by using
the Nambu-Jona-Lasinio (NJL) model~\cite{Toublan:2003tt,
Barducci:2004,L.-y.He:2005,L.-y.He:2007,L.-y.He:2006,Andersen:2009},
the random matrix model~\cite{Klein:2003fy},
the strong coupling lattice analysis~\cite{Nishida:2004},
Ginzburg-Landau approach~\cite{Abuki:2013vwa},
the hadron resonance gas model~\cite{Toublan:2004ks},
holographic QCD models~\cite{Kim:2007gq,Kim:2007xi,Parnachev:2007bc,Aharony:2007uu,
Basu:2008bh,Ammon:2009fe,Albrecht:2010eg,
Rust:2010gq,Kim:2011gw,Kim:2012zzk,Lee:2013oya}.

Although there are so many works on the pion condensation at finite isospin chemical potential,
there are 
not many works for studying the strength of the chiral symmetry breaking.
Namely, it is interesting to ask whether or not the chiral symmetry is partially restored in the isospin matter.

In Refs.~\cite{L.-y.He:2005,L.-y.He:2007,L.-y.He:2006,Andersen:2009}, based on the NJL model analysis, the authors seemed to conclude that the reduction of $\langle\bar{q} q\rangle$ in the isospin matter implies the partial chiral symmetry restoration.
In Ref.~\cite{Abuki:2013vwa},
the Ginzburg-Landau approach is used to study the the $\langle \bar{q} q \rangle$ condensate together with the pion condensation.
However, the absolute strength of the chiral symmetry breaking,
which is characterized by the chiral condensate 
$\tilde{\sigma} = \sqrt{ \langle \sigma \rangle^2 + \langle \pi^a \rangle^2 }$,
 is not clearly studied.
The analysis using the strong coupling lattice in Ref.~\cite{Nishida:2004} 
shows that $\tilde{\sigma}$ decreases in the high isospin chemical potential associated with the decreasing pion condensation. 
The decreasing pion condensation might be a special feature in the strong coupling lattice analysis, so that it would be interesting to
study the behavior of $\tilde{\sigma}$ using the various ways.

In this work, we study
the pion condensation phase in a holographic QCD model~\cite{EKSS,RP1}
by solving the equations of motion for mean fields corresponding to $\pi$, $\sigma$ and the time component of
$\rho$ meson.
Our results show that
the phase transition is of the second order consistently with the one obtained in the $O(p^2)$ chiral 
Lagrangian~\cite{SS}, while it is contrary to the result in Ref.~\cite{Albrecht:2010eg}.
It is remarkable that 
the chiral condensate defined by
$\tilde{\sigma} \equiv \sqrt{ \langle \sigma \rangle^2 + \langle \pi^a \rangle^2 }$
is almost constant in the small $\mu_I$ region, while it grows
with $\mu_I$ in the large $\mu_I$ region.
This implies that 
the chiral symmetry breaking is enhanced
by the existence of the isospin chemical potential.

This paper is organized as follows:
In section~\ref{sec:model} we introduce some basic point of the model which we use in the present analysis.
Section~\ref{sec:PCP}
is devoted to the main part of this paper, where we study the pion condensation together with the chiral condensation.
In section~\ref{sec:chiral}
we analyze our result in terms of the chiral Lagrangian at $O(p^4)$.
Finally, we give a summary and discussions in section~\ref{sec:summary}

\section{Model}
\label{sec:model}

In the present analysis we use 
the hard-wall holographic QCD model given in Refs.\cite{EKSS,RP1}.
The action is given by
\begin{equation}
S_5=\int d^4x \int_\epsilon^{z_m} dz\mathcal{L}_5
\ ,
\label{action}
\end{equation}
where $\epsilon$ and $z_m$ are the UV and the IR cutoffs. 
The 5-dimensional Lagrangian is
\begin{equation}
\mathcal{L}_5
=\sqrt{g} \mathrm{Tr} \left[ |DX|^2 +3|X|^2 -\frac{1}{4g_5^2}\left(F^2_L+F^2_R\right)\right]
+\mathcal{L}^{BD}_5,
\label{Lagrangian1}
\end{equation}
where the metric is given by
\begin{equation}
ds^2=a^2(z) \left(\eta_{\mu\nu}dx^\mu dx^\nu -dz^2\right)
\label{metric}
\end{equation}
with $a(z)=1/z$. 
The covariant derivative and the field strength are given by 
\begin{eqnarray}
D_M X & =& \partial_M X-iL_M X + i X R_M
\ ,
\label{D}
\\
F^L_{MN}&=&\partial_M L_N -\partial_N L_M - i\left[ L_M , L_N \right]\ .
\label{strength}
\end{eqnarray}
where
$M=(\mu,5)$ is the 5th dimensional indices.
$\mathcal{L}^{BD}_5$ in Eq.~(\ref{Lagrangian1})
is the boundary term introduced as~\cite{RP2}
\begin{equation}
\mathcal{L}^{BD}_5=-\sqrt{g} \mathrm{Tr} \left\{ \lambda z_m |X|^4-m^2 z_m |X|^2 \right\} \delta (z-z_m) \ ,
\label{Lagrangian BD}
\end{equation}
where $z_m$ in the coefficients of the $\vert X\vert^4$ term and the $\vert X \vert^2$
term are introduced in such a way that $\lambda$  and $m^2$ carry no dimension. 
This model has a chiral symmetry corresponding to U(2$)_{\rm L}\times$U(2)$_{\rm R}$. 
There exists the Chern-Simons term in addition to the above term.
However, the CS term does not contribute to the pion condensation when the spatial rotational symmetry is assumed as in this paper.

The scalar field $X$ and the gauge fields $L_M$ and $R_M$ transform
under the  U$(2)_L\times $U$(2)_R$ as 
\begin{eqnarray}
X &\rightarrow & X'= g_L X g_R^\dagger
\ ,
\label{X trans}\\
L_M &\rightarrow& {L'}_M = g_L L_M g_L^\dagger -i g_L^\dagger \partial_M g_L
\ ,
\label{L trans}
\end{eqnarray}
where $g_{L,R} \in \mbox{U}(2)_{L,R}$
are the transformation matrices of the chiral U$(2)_L\times $U$(2)_R$
symmetry.
In the following analysis we adopt the $L_5=R_5=0$ gauge, and use the IR-boundary 
condition $F^L_{5\mu}|_{z_m}=F^R_{5\mu}|_{z_m}=0$.

In the vacuum 
the
chiral symmetry is spontaneously broken down to U$(2)_V$ by the vacuum expectation value of $X$. 
This is given by solving the equation of motion as~\cite{EKSS,RP1}
\begin{equation}
X_0(z)=\frac{1}{2}\left(m_q z + \sigma z^3 \right)\equiv \frac{1}{2} {v}(z) 
\ ,
\label{VEV}
\end{equation}
where $m_q$ corresponds to the current quark mass and $\sigma$ to 
the quark condensate.
They are related with each other by the IR-boundary condition:
\begin{equation}
z_m \partial_5 {v} |_{\rm IR}=
\left. -\frac{{v}}{2} \left(\lambda {v}^2 - 2 m^2 \right) \right\vert_{\rm IR}
\ .
\label{IR for v}
\end{equation}
The fields are parameterized as
\begin{eqnarray}
X&=&\frac{1}{2}\left(S^0 \sigma^0 +S^a \sigma^a  \right)e^{i\pi^b \sigma^b+i \eta}
\ ,
\label{parameterize1}\\
V_\mu&=&\frac{L_\mu + R_\mu }{2}
\ ,
\label{parameterize2}\\
A_\mu&=&\frac{L_\mu - R_\mu }{2}
\ ,
\label{parameterize3}\\
V^A_\mu&=&\mathrm{Tr}\left[V_\mu \;\sigma^A \right]
\ ,
\\
A^A_\mu&=&\mathrm{Tr}\left[A_\mu\;\sigma^A \right]
\end{eqnarray}
where  $\sigma^a$ ($a=1,2,3$) are Pauli matrices and  $\sigma^0=1$, and
the superscript index $A$ runs over 0, 1, 2 and 3.
The iso-singlet scalar part $S^0$ is separated into a background field part and a fluctuation part as  $S^0={v}+\tilde{S}^0$.
A parameter $g_5^2$ is determined by matching with QCD as
\begin{eqnarray}
g_5^2=\frac{12\pi^2}{N_c}.
\label{parameter}
\end{eqnarray}
The pion is described as a linear combination of the lowest eigenstate of 
$\pi^a$ and the longitudinal mode of $A^a_\mu$,  and the $\rho$ meson is the lowest eigenstate of $V_\mu$. 
The values of the $m_q$ and $z_m$ 
together with that of $\sigma$ are fixed by fitting them to the pion mass  $m_\pi=139.6$ MeV, the $\rho$ meson mass $m_\rho=775.8$ MeV and the pion decay constant $f_\pi=92.4$ MeV as~\cite{EKSS,RP1}
\begin{equation}
m_q = 2.29\,\mbox{MeV}\ , \quad
z_m = 1/(323\,\mbox{MeV}) \ , \quad
\sigma = \left( 327 \,\mbox{MeV}\right)^3
\ .
\label{parameter1}
\end{equation}
By using this value of $\sigma$ and a scalar meson mass as inputs,
the values of the parameters $m^2$ and $\lambda$ in the boundary potential
are fixed.~\cite{RP2}
It was shown~\cite{RP2} that there is an upper bound for the scalar meson mass
as $1.2$\,GeV, 
but the dependence on the mass on the value of $\lambda$ is small.
So in the present analysis, we use the $a_0$ meson mass $m_{a_0}=980$ MeV
as a reference value, which fixes $m^2 = 5.39$ 
and $\lambda=4.4$, and see the dependence of our results on the scalar
meson mass.

\section{Pion condensation phase}
\label{sec:PCP}

In this section we study the pion condensation for finite isospin chemical potential $\mu_I$
in the holographic QCD model explained in the previous section.
Since the pion mass exists in the present model. We will have a phase transition
from the normal phase to the pion condensation phase for increasing $\mu_I$.
In the present paper we are interested in the pion condensation phase 
for small isospin chemical potential, so that we assume that the rotational 
symmetry O(3) is not broken by e.g. the $\rho$ meson condensation.
 We also assume the time-independent condensate, 
then the vacuum structure is determined by studying the mean fields
of five-dimensional fields which do not depend on the four-dimensional coordinate.

We introduce the isospin chemical potential $\mu_I$ 
as a UV-boundary value of the time component of 
the gauge field of SU(2)$_{\rm V}$ symmetry as
\begin{eqnarray}
V^3_0(z)|_\epsilon=\mu_I \ ,
\label{mu}
\end{eqnarray}
where the superscript $3$ indicates the third component of the isospin corresponding to the neutral $\rho$ meson.
The  assumption of rotational invariance implies
 $L_i=R_i=0$. 
Then the wave functions of 
$V^a_0(z)$ ($a=1,2,3$),  $\pi^a_0(z)$ and $A^a_0(z)$ 
are determined by solving the equations of motion.

In the present analysis,
the CS term, given as 
\begin{align}
S_{\rm {CS}}\propto &\int d^4x \int_\epsilon^{z_m} dz
\nonumber \\&
\times \epsilon^{MNPQS} L^0_M
 \mathrm{Tr}\left[F^L_{NP}F^L_{QS}
\right]
 -({\rm{L}} \rightarrow {\rm{R}})
\ ,
\end{align}
vanishes because this 
term must proportional to $L_i^a$ or $R^a_i$, where $L_M^0$ is a gauge field  corresponding to U$(1)_L$ i.e. 
$L_M^0=\mathrm{Tr}\left[L_M\right]$.  
The gauge field corresponding to U$(1)_V$, which includes the $\omega$ meson
and its radial excitations, does not show up in our analysis, 
because it couples to other fields only through
 the CS term.

The $X$ field consists of eight degrees of freedom, which include $\eta$ and $S^a$ ($a=1,2,3$). 
Since 
the $\eta$ is isosinglet, 
it does not condense by itself.
However, the existence of  $S^a$ condensation ($a_0$ meson condensation)
together with the pion condensation 
triggers the $\eta$ condensation.
The $a_0$ meson condensation will occur for $\mu_I \ge m(a_0)$.
 Since we study the region of $\mu_I \leq m_\rho$, we expect that 
both $\eta$ and $a_0$ condensations vanish in this region.
Actually, we can check that $\eta= S^a=0$ is a solution of the equations of motion
in the following way:
By using the parametrization of Eq.~(\ref{parameterize1}), the terms including $\eta$ and $S^a$of Lagrangian (\ref{Lagrangian1}) are written as
\begin{align}
\mathcal{L}_5
~\sim&~~
\frac{a^3}{4}\mathrm{Tr}\left[SS\left\{
\left(U\partial_5 U^\dagger\right)^2+ L_0L_0+ R_0R_0
+a^2\right\}
\right.\nonumber \\&
-(\left(S^0\right)^2+SS)(\partial_5 \eta)^2
+4iS^0(\partial_5 \eta)S
\left(U\partial_5 U^\dagger\right)
\nonumber\\&
-\left[S,\left(\partial_5 S\right)\right]\left(U\partial_5 U^\dagger\right)
-\left(\partial_5 S\right)^2 
- 2SL_0SUR_0U^\dagger 
\nonumber\\&\left.
+8S^0 A^0_0  S \left(L_0-UR_0U^\dagger\right) \right]
\label{eta and S} 
\end{align}
where $S=S^a \sigma^a$ and $U=e^{i\pi^a \sigma^a}$.
It is easy to confirm that $\eta = S^a =0$ together with $A^0_0=0$ is 
a solution of the equations of motion 
for them.
Then, in following analysis, 
we take $\eta=S^a=0$.

For writing the equations of motion for mean fields,
it is convenient to express
\begin{equation}
e^{i\pi^a \sigma^a}= \cos b \;1 + i \sin b \left( n^a \sigma^a\right)\ ,
\end{equation}
where
\begin{equation}
b=\sqrt{\pi^b \pi^b}\ , \quad
n^a=\frac{\pi^a}{b} \ .
\end{equation}
We include the condition 
$(n^a)^2=1$ into the Lagrangian using a Lagrange multiplier
$\tilde{\lambda}$.
\begin{widetext}
Now, the coupled equations of motion are given as
\begin{eqnarray}
\partial_5 \left( - a^3 \left(S^0\right)^2 \sin^2 b\partial_5 n^3 \right) 
&-a^3\left(S^0\right)^2&
\left[\sin^2 b \;n^3 ( n^b V^b_0 ) 
+ \sin^2 b ~A_0^3\left( n^b A^b_0 \right)
+ \sin b\cos b \; \epsilon^{b3c} A^b_0 V^c_0 \right] +2 \tilde{\lambda} n^3=0,
\label{EOM n3}
\nonumber \\ 
\partial_5 \left( \frac{a}{g^2_5} \partial_5 A_0^3 \right)
&-a^3\left(S^0\right)^2&
\left[ \cos^2 b \; A_0^3+ \sin^2 b \;n^3\left( n^b A^b_0 \right)
-\sin b \cos b \; \epsilon^{bc3}  V^b_0 n^c\right]=0,
\label{EOM A}
\nonumber \\ 
\partial_5 \left( \frac{a}{g^2_5} \partial_5 V_0^1 \right)
&-a^3\left(S^0\right)^2&
\left[\sin^2 b\; V^1_0 - \sin^2 b\; n^1 \left(n^b V^b_0 \right) 
+ \sin b \cos b \; \epsilon^{bc1} A_0^b n^c \right]=0,
\label{EOM V1}
\nonumber \\ 
\partial_5 \left( \frac{a}{g^2_5} \partial_5 V_0^2 \right)
&-a^3\left(S^0\right)^2&
\left[\sin^2 b\; V^2_0 - \sin^2 b\; n^2 \left(n^b V^b_0 \right) 
+ \sin b \cos b \; \epsilon^{bc2} A_0^b n^c \right]=0\ ,
\label{EOM V2}
\end{eqnarray}
where the summation over the indices $b$ and $c$ are understood.
\end{widetext}
We can easily check that $\pi^3=A^3_0=0$ together with $V^1_0=V^2_0=0$ gives a set of solutions 
for the above coupled equations of motion, which implies that the neutral pion does 
not condense. Then, in the following we assume that this set of solution is 
physically realized, and take $\pi^3=A^3_0=0$ and $V_0^1=V_0^2=0$.

At finite isospin chemical potential this  theory has 
the U(1) symmetry 
which is a subgroup of the isospin SU(2)$_V$.
Using this U(1) symmetry, we rotate away the condensation of  the $\pi^2$ field 
to keep only the $\pi^1$ condensation.
\begin{widetext}
By setting $\pi^3=A^3_0=0$ and $V_0^1=V_0^2=0$ and writing 
\begin{eqnarray*}
e^{i\pi^a \sigma^a} &=&
\cos b \;1 + i \sin b \; \sigma^1 ,
\nonumber\\
A^a_0 &=&\theta~(\cos \zeta ,\sin \zeta ,0)\ ,
\nonumber\\
V^3_0 &=&\varphi + \mu_I \ ,
\end{eqnarray*}
the Lagrangian is rewritten as
\begin{eqnarray}
\mathcal{L}_5
&=&
\frac{a^3}{2}\left[-\left(\partial_5 S^0\right) ^2
-\left(S^0\right)^2\left(\partial_5 b\right)^2
\right]\nonumber \\ &&
{} +\frac{a^3\left(S^0\right)^2}{2}\left[\sin^2 b ~( \varphi +\mu_I )^2
			-\theta\sin 2b \sin \zeta ~(\varphi +\mu_I) 
		+\theta^2- \theta^2 \sin^2 b \sin^2 \zeta\right]
\nonumber \\ &&
                 {} +\frac{3a^5}{2} \left(S^0\right)^2
                  {} +\frac{a}{2g_5^2}\left[\left(\partial_5 \varphi \right)^2 
	+ \left(\partial_5 \theta \right)^2
		+\theta^2 \left(\partial_5 \zeta \right)^2 \right].
\label{Lagrangian2}
\end{eqnarray}
{}From the above Lagrangian,
the equations of motion are obtained as
\begin{eqnarray} &&
\partial_5 \left( -a^3 \partial_5 S^0 \right) + a^3 S^0 \left(\partial_5 b\right)^2
 -3a^5S^0
\nonumber \\&&
~~~~~~~~~~-a^3S^0\left[\sin^2 b \; ( \varphi +\mu_I )^2
-\theta\sin 2b \;\sin \zeta \; (\varphi +\mu_I) 
+\theta^2 - \theta^2 \sin^2 b \; \sin^2 \zeta \right] =0,
\nonumber \\ &&
\partial_5 \left( - a^3 \left(S^0\right)^2 \partial_5 b \right) 
-\frac{a^3\left(S^0\right)^2}{2}\left[\sin 2b \; \left\{( \varphi +\mu_I )^2
- \theta^2 \sin^2 \zeta\right\}-2\theta \cos 2b \;\sin \zeta 
~( \varphi +\mu_I )\right]=0,
\nonumber \\ &&
\partial_5\left(\frac{a}{g_5^2}\partial_5 \theta \right)
-\frac{a}{g_5^2}\theta \left( \partial_5 \zeta \right)^2
-\frac{a^3\left(S^0\right)^2}{2}\left[-\sin 2b \; \sin \zeta 
~( \varphi +\mu_I )+ 2\theta \left\{ 1 -  \sin^2 b \;\sin^2 \zeta \right\}\right]=0,
\nonumber \\ &&
\partial_5\left(\frac{a}{g_5^2}\theta^2\partial_5 \zeta \right)
+\frac{a^3\left(S^0\right)^2}{2}\left[-\theta \sin 2b \;\cos \zeta \;\;( \varphi +\mu_I )
- \theta^2 \sin 2 \zeta\right]=0,
\nonumber \\ &&
\partial_5 \left( \frac{a}{g^2_5} \partial_5 \varphi \right)
-\frac{a^3\left(S^0\right)^2}{2}\left[2\sin^2 b 
\;( \varphi +\mu_I )-\theta\sin 2b \;\sin \zeta \right]=0.
\label{EOM}
\end{eqnarray}
\end{widetext}
Using the
boundary conditions  listed in Table~\ref{Boundary conditions},
we solve the above coupled  
equations of motion to  determine the isospin chemical potential as an eigenvalue. 
Then, we calculate the isospin number density from the
following formula obtained 
from the Lagrangian (\ref{Lagrangian2}):
\begin{eqnarray}
n_I&=&\int dz \frac{\partial \mathcal{L}_5}{\partial \mu_I}
\nonumber \\
&=&
\int dz \frac{a^3\left(S^0\right)^2}{2}\left[2\sin^2 b ( \varphi +\mu_I )-\theta\sin 2b \sin \zeta \right]\ .
\nonumber\\
\label{iso-density}
\end{eqnarray}
\begin{table}[h]
 \begin{center}
 \begin{tabular}{|c|c|c|}\hline
variables	&	UV	&	IR	
\\\hline
$S^0$	&	$\frac{S^0}{z} |_\epsilon=m_q$	&	
$ \partial_5 S^0 |_{z_m}=
\left. -\frac{S^0}{2z_m} \left(\lambda \left(S^0\right)^2 - 2 m^2 \right) \right\vert_{z_m}$
\\\hline
$b$	&	$b |_\epsilon=0$	&	$\partial_5 b |_{z_m}=0$
\\\hline
$\theta$	&	$\theta |_\epsilon=0$	&	$\partial_5 \theta |_{z_m}=0$
\\\hline
$\zeta$	&	$\zeta |_\epsilon=\frac{\pi}{2}$	&	$\partial_5 \zeta |_{z_m}=0$
\\\hline
$\varphi$	&	$\varphi |_\epsilon=0$	&	$\partial_5 \varphi |_{z_m}=0$
\\\hline
 \end{tabular}
 \end{center}
\caption{Boundary conditions of variables.}
\label{Boundary conditions}
\end{table}

\begin{figure}[ht]
 \begin{center}
  \includegraphics[width=70mm]{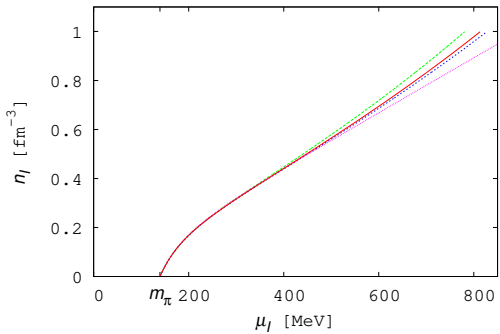}
 \end{center}
 \caption[]{
Relation between the
isospin number density $n_I$ and the isospin number chemical potential $\mu_I$.
The green, red and blue  curves show our results for $\lambda = 1$, $4.4$ and $100$,
respectively.
The pink dashed-curve shows the result given by the chiral Lagrangian in Ref.~\cite{SS}. 
Each choice of $\lambda$ corresponds to $m_{a_0}=610$MeV, $980$MeV
and $1210$MeV, respectively.
}
 \label{fig nI vs mu}
\end{figure}

We show the resultant relation between the isospin chemical potential and the 
isospin density in Fig.~\ref{fig nI vs mu}
for $\lambda = 1$, $4.4$ and $100$ corresponding to
$m_{a_0}=610$MeV, $980$MeV
and $1210$MeV.
This shows that the phase transition is of the second 
order and the critical chemical potential is predicted to be equal to 
the pion mass.
This is consistent with the result obtained by the chiral Lagrangian approach
in Ref.~\cite{SS}, but contrary to the result in Ref.~\cite{Albrecht:2010eg}. 
Furthermore,
our result on 
the relation between isospin number density and isospin chemical potential 
for small $\mu_I$ 
agrees with the following one obtained by O$(p^2)$ chiral Lagrangian~\cite{SS}:
\begin{eqnarray}
n_I&=&f_\pi^2 \mu_I\left(1-\frac{m_\pi^4}{\mu_I^4}\right).
\label{iso-density SS}
\end{eqnarray}
For $\mu_I > 500$\,MeV, there is a difference between our predictions
and the one from O$(p^2)$ chiral Lagrangian, which can be understood as
the higher order contribution as we will show in the next section.

\begin{figure}[ht]
 \begin{center}
 	 \includegraphics[width=70mm]{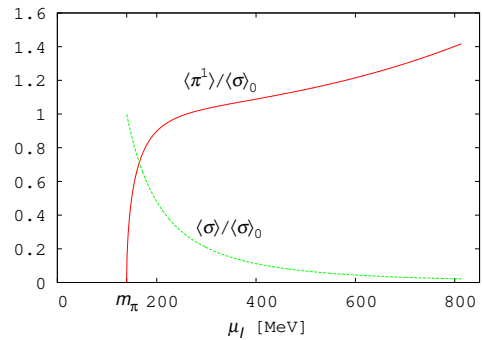}
 \end{center}
 \caption{$\mu_I$ dependences of the $\pi$-condensate (red curve) and
the ``$\sigma$"-condensate (green curve). }
 \label{fig v vs mu}
\end{figure}

We next study the $\mu_I$ dependences of the ``$\sigma$"-condensate
corresponding to $\left\langle \bar{q} q \right\rangle$ and the $\pi$-condensate
to 
$\langle \bar{q} \gamma_5 \sigma^a q \rangle$.
For obtaining these condensate through 
the AdS/CFT correspondence, 
we introduce the scalar source $s$ and the pesudoscalar source $p^a$ as
\begin{eqnarray}
\frac{\delta X}{z}&=&\frac{\delta X^\dagger}{z}=\frac{s}{2}1,
\nonumber \\
\frac{\delta X}{z}&=&-\frac{\delta X^\dagger}{z}=\frac{ip^a}{2}\sigma^a.
\end{eqnarray}
The UV-boundary term of $X$ is written as
\begin{eqnarray}
\delta S^{UV}&=&
\int d^4x ~\mathrm{Tr} \left[\frac{\delta X^\dagger}{z} a^2\left(\partial_5 X\right)
+h.c.\right]_\epsilon 
\nonumber \\
&=&
\int d^4x ~\mathrm{Tr} \left[\frac{\delta X^\dagger}{z} 
a\left(\partial_5 \frac{X}{z} + \frac{X}{z^2}\right)
+h.c.\right]_\epsilon .
\nonumber \\
\label{UV of X}
\end{eqnarray}
We neglect the second term in the last line of the above equation.
Then the $\pi$-condensate and ``$\sigma$''-condensate are defined 
as~\footnote{
Note that we use
$ \langle\bar{q}q\rangle = \langle\bar{u}u\rangle+\langle\bar{d}d\rangle$.
}
\begin{align}
\langle \bar{q} \gamma_5 \sigma^a q \rangle
\equiv&
\frac{1}{2}\mathrm{Tr} \left[i\sigma^a a\left(\partial_5 \frac{X}{z}\right)
+h.c.\right]_\epsilon 
=\langle\pi^a\rangle,
\nonumber \\
\langle \bar{q} q \rangle \equiv& 
\frac{1}{2}\mathrm{Tr} \left[ a\left(\partial_5 \frac{X}{z}\right)
+h.c.\right]_\epsilon 
=\langle\sigma\rangle
\ .
\end{align}
We show the $\mu_I$ dependences of these condensate in Fig.~\ref{fig v vs mu},
where 
$\langle\sigma\rangle_0$ is the ``$\sigma$"-condensate at $\mu_I=0$.
This shows that 
the ``$\sigma$''-condensate decreases rapidly after the phase transition where the $\pi$-condensate grows rapidly.
The ``$\sigma$"-condensate becomes very small for $\mu_I \gtrsim 400$\,MeV, 
while the 
$\pi$-condensate keeps increasing.
Using the form $\langle\pi^a\rangle \propto \left(\mu_I-\mu_I^c\right)^\nu$ near the
phase transition point, we fit 
the critical exponent $\nu$ to obtain 
$\nu=\frac{1}{2}$.
This implies that 
the phase transition here is the mean field type.

\begin{figure}[ht]
 \begin{center}
   \includegraphics[width=70mm]{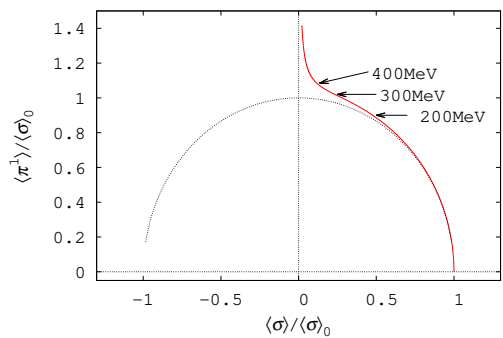}
 \end{center}
 \caption{The chiral circle is showed as the red curve.  The black curve is an unit circle.}
 \label{fig chiral circle}
\end{figure}

We also show
the ``chiral circle"  in 
Fig.~\ref{fig chiral circle}. 
It is remarkable that
the value of the ``chiral condensate" defined by
\begin{equation}
\tilde{\langle \sigma\rangle} = \sqrt{ \langle \sigma \rangle^2 + \langle \pi^a \rangle^2 }
\end{equation}
is constant for increasing isospin chemical potential $\mu_I$ for $\mu_I \lesssim 300$\,MeV,
and that it grows rapidly in the large $\mu_I$ region.

\section{Comparison with the chiral Lagrangian}
\label{sec:chiral}

In this section, we compare our prediction on the relation between the
isospin number density and the isospin chemical potential shown in 
Fig.~\ref{fig nI vs mu} as well as the $\mu_I$-dependences of the $\pi$-condensate and the ``$\sigma$''-condensate in Fig.~\ref{fig v vs mu}, 
with the ones from the chiral Lagrangian including the O($p^4$) terms.
Here we use the following 
chiral Lagrangian for two flavor case~\cite{GL,Harada:2003jx}:
\begin{eqnarray}
\mathcal{L}^{\rm {ChPT}}
&=&
\frac{1}{4}F_0^2\mathrm{Tr}\left[D_\mu U D^\mu U^\dagger\right]
+\frac{1}{4}F_0^2\mathrm{Tr}\left[\chi^\dagger U +U^\dagger \chi\right]
\nonumber \\ &&
+L_1^{(2)}\left(\mathrm{Tr}\left[D^\mu U^\dagger D_\mu U\right]\right)^2
\nonumber\\ &&
+L_2^{(2)}\mathrm{Tr}\left[D_\mu U^\dagger D_\nu U\right]
\mathrm{Tr}\left[D^\mu U^\dagger D^\nu U\right]
\nonumber \\ &&
+L_4 ^{(2)}\mathrm{Tr}\left[D^\mu U^\dagger D_\mu U \right]
\mathrm{Tr}\left[ \chi^\dagger U +U^\dagger \chi\right]
\nonumber \\ &&
+L_6^{(2)} \left(\mathrm{Tr}\left[ \chi^\dagger U +U^\dagger \chi\right]\right)^2
\nonumber\\ &&
+L_7^{(2)} \left(\mathrm{Tr}\left[ \chi^\dagger U -U^\dagger \chi\right]\right)^2
\nonumber \\ &&
+L_8^{(2)} \mathrm{Tr}\left[ \chi^\dagger U \chi^\dagger U +\chi U^\dagger \chi U^\dagger\right]
\nonumber \\ &&
+iL_9^{(2)} \mathrm{Tr}\left[F^{R}_{\mu\nu} D^\mu U^\dagger D^\nu U 
+F^{L}_{\mu\nu} D_\mu U D^\mu U^\dagger \right]
\nonumber\\ &&
+L_{10}^{(2)} \mathrm{Tr}\left[U^\dagger F^{L}_{\mu\nu} U F^{R\mu\nu} \right]
\nonumber\\ &&
+H_1^{(2)}\mathrm{Tr}\left[ F^{L\mu\nu}F^{L}_{\mu\nu} +F^{R\mu\nu}F^{R}_{\mu\nu} \right]
\nonumber\\ &&
+H_2^{(2)}\mathrm{Tr}\left[\chi^\dagger \chi \right]
\ ,
\label{Lagrangian-chiral}
\end{eqnarray}
where $U$ is parametrized by the pseudoscalar meson fields
as
\begin{equation}
U = e^{i 2 \pi/f_\pi } \ , \quad \pi = \pi^a T_a \ .
\end{equation}
$\chi$ includes the scalar and pseudoscalar source fields, and
the covariant derivative $D_\mu U$ is expressed 
as
\begin{equation}
D_\mu U \equiv \partial_\mu U - i {\mathcal L}_\mu U + i U {\mathcal R}_\mu
\ , 
\end{equation}
 where 
${\mathcal L}_\mu$ and ${\mathcal R}_\mu$ are external gauge fields corresponding to 
SU$(2)_{L,R}$.

In the following analysis, we will study the relation between the isospin number density and 
the isospin chemical potential as well as the condensates using the above chiral Lagrangian at tree level.
In the ordinary chiral perturbation theory, the tree-level contribution from O($p^4$) terms 
are of the same order as the one-loop contribution of O($p^2$), so that both contributions 
should be included.
 However, because one-loop corrections 
are counted as next to leading order in $1/N_c$ expansion, we neglect the one-loop corrections
in the present analysis.
Then, one can simply introduce
the isospin chemical potential $\mu_I$ as vacuum expectation values of 
these external gauge fields as
$\langle {\mathcal L}^3_\mu \rangle = \langle {\mathcal R}^3_\mu \rangle 
= \frac{\mu_I}{2} \delta_{0\mu}$.
In this case, $\mu_I$ appears only through the covariant derivative as
\begin{eqnarray}
D_0 U =\partial_0 U -i\frac{\mu_I}{2}[\sigma^3 , U ]
\label{chemical-potential-chiral}
\end{eqnarray}
where $\sigma^3$ is the third component of the Pauli matrices. 
Parameterizing $U$ 
as 
$U=\cos \alpha 
+ i\sin \alpha \; \left( \sigma^1 \cos \phi + \sigma^2 \sin \phi \right)$,
we get the effective potential as
\begin{eqnarray}
V_{\scriptsize {\mbox {eff}}}&=&-\mathcal{L}^{\rm {ChPT}}
 \nonumber \\
&=&
-\frac{f_\pi^2}{2}\mu_I^2\left(2-\beta\right)\beta
-f_\pi^2m_\pi^2\left(1-\beta\right)
\nonumber \\ &&
-4A\mu_I^4\left(2-\beta\right)^2\beta^2
+8Cm_\pi^2\mu_I^2\left(2-\beta\right)\beta^2
\nonumber \\ &&
-8B m_\pi^4\beta^2+(\mbox{const.})
\label{effective-potential}
\end{eqnarray}
where $\beta \equiv 1 - \cos \alpha$
and 
\begin{eqnarray}
A&\equiv& L^{(2)}_1 + L^{(2)}_2 \ , \nonumber\\
B&\equiv& 2L^{(2)}_6 + L^{(2)}_8 \ , \nonumber\\
C&\equiv& L^{(2)}_4 \ .
\label{ABC} 
\end{eqnarray}
We determine the value of $\beta$ by minimizing the above $V_{\rm eff}$, and then calculate the isospin number density, the ``$\sigma$"-condensate and 
the $\pi$-condensate through
\begin{align}
n_I=&-\frac{\partial V_{\rm {eff}}}{\partial \mu_I}
\nonumber \\ 
=&
f_\pi^2\mu_I\left(2-\beta\right)\beta
+16A\mu_I^3\left(2-\beta\right)^2\beta^2
\nonumber \\ &
-16Cm_\pi^2\mu_I\left(2-\beta\right)\beta^2
\ ,
\label{density-chiral}
\\
\frac{\langle \sigma \rangle}{\langle \sigma \rangle_0}
=&
1-
\kappa\left[1
-\left(8C\frac{\mu_I^2}{f_\pi^2} (2-\beta)-16B \frac{m_\pi^2}{f_\pi^2}\right)(1-\beta)\right]\beta
\ ,
\\
\frac{\langle \pi \rangle}{\langle \sigma \rangle_0}
=&\kappa
\left[1+\left(8C\frac{\mu_I^2}{f_\pi^2}(2-\beta)
-16B \frac{m_\pi^2}{f_\pi^2}\right)\beta \right]\sqrt{\beta(2-\beta)}
\ ,
\end{align}
where
\begin{align}
\kappa=&\frac{f_\pi^2m_\pi^2}{m_q \langle \sigma \rangle_0}.
\end{align}
 We set $\kappa \simeq 1.04$~\footnote{
The deviation of $\kappa$ from $1$ expresses the deviation from the Gell-Mann-Oakes-Renner relation  due to the contribution of $H^{(2)}_2-2L^{(2)}_8$.
}, which is given by using $m_\pi=139.6$ MeV, $f_\pi=92.4$ MeV and (\ref{parameter1}).

\begin{figure}[ht]
 \begin{center}
   \includegraphics[width=70mm]{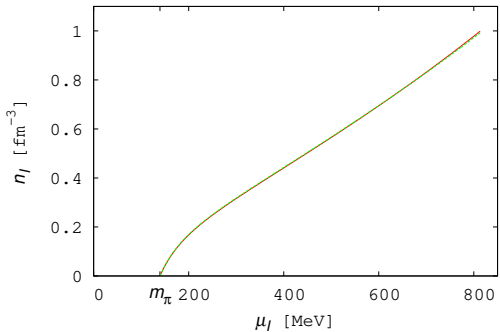}
 \end{center}
 \caption{
$\mu_I$ dependence of $n_I$ obtained from the
$O(p^4)$ chiral Lagrangian for the best fitted values of $A$, $B$ and $C$ (green curver)
compared with our result (red curve).
}
 \label{fig fitting}
\end{figure}

\begin{figure}[ht]
 \begin{center}
   \includegraphics[width=70mm]{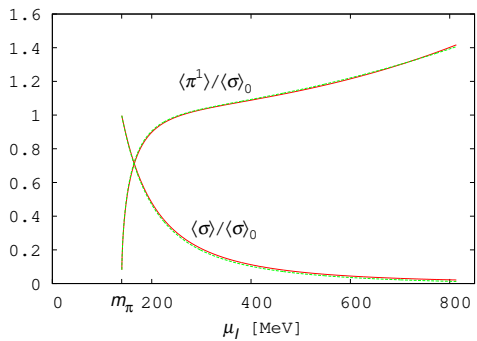}
 \end{center}
 \caption{
$\mu_I$ dependence of $\frac{\langle \sigma \rangle}{\langle \sigma \rangle_0}$ and 
$\frac{\langle \pi^1 \rangle}{\langle \sigma \rangle_0}$ obtained from the
$O(p^4)$ chiral Lagrangian for the best fitted values of $A$, $B$ and $C$ (green curver)
compared with our result (red curve).
}
 \label{fig fitting2}
\end{figure}

We fit the values of 
the coefficients $A$, $B$ and $C$ to our result on the $\mu_I$ dependence of $n_I$
, $\frac{\langle \sigma \rangle}{\langle \sigma \rangle_0}$ and 
$\frac{\langle \pi^1 \rangle}{\langle \sigma \rangle_0}$
shown in Figs.~\ref{fig nI vs mu} and \ref{fig v vs mu},
by  minimizing 
\begin{align}
\sum^{}_{\rm {date}}&\left[
\left(
\left.\frac{n_I}{f_\pi^2\mu_I}\right|_{\rm {result}} -
\left.\frac{n_I(A,B,C)}{f_\pi^2\mu_I}\right|_{\rm {ChPT}}
\right)^2
\right.
\nonumber\\
+&\left(
\left.
\frac{\langle \sigma \rangle}{\langle \sigma \rangle_0}
\right|_{\rm {result}}
- \left.
\frac{\langle \sigma \rangle}{\langle \sigma \rangle_0}
\right|_{\rm {ChPT}}
\right)^2
\nonumber\\
+&\left.
\left(
\left.
\frac{\langle \pi^1 \rangle}{\langle \sigma \rangle_0}
\right|_{\rm {result}} 
-\left.
\frac{\langle \pi \rangle}{\langle \sigma \rangle_0}
\right|_{\rm {ChPT}}
\right)^2
\right]
\ .
\end{align}
We show the best fitted values of $A$, $B$ and $C$ for $\lambda = 4.4$ in Table~\ref{coefficients}
together with a set of empirical 
values.~\footnote{
We calculate the empirical values of $A$, $B$ and $C$ through Eq.~(\ref{ABC}),
where $L_i^{(2)}$ are determined from the values of low energy constant 
in the two flavor
ChPT~\cite{Gasser:1983yg} 
with the renormalization scale equal to $M_\eta$.
}
\begin{table}[h]
 \begin{center}
 \begin{tabular}{|c|c|c|c|}\hline
	&$A\times 10^3$	&	$B\times 10^3$	&	$C\times 10^3$	
\\\hline
Fitting result	&$0.093$&$1.01$&$0.63$
\\\hline
Empirical values&$0.4\pm 1.2$&$1.2\pm 0.9$&$1.1\pm 0.6$
\\\hline
 \end{tabular}
 \end{center}
\caption{
The best fitted values of 
$A\equiv L^{(2)}_1 + L^{(2)}_2$, $B\equiv 2L^{(2)}_6 + L^{(2)}_8$ and 
$C\equiv L^{(2)}_4$
compared with a set of empirical values.
}
\label{coefficients}
\end{table}
We show the $\mu_I$ dependence of $n_I$ in 
Fig.~\ref{fig fitting} and $\mu_I$ dependence of $\frac{\langle \sigma \rangle}{\langle \sigma \rangle_0}$ and 
$\frac{\langle \pi^1 \rangle}{\langle \sigma \rangle_0}$ in 
Fig.~\ref{fig fitting2}.
These figures show that the deviation of our result from the one 
obtained from the $O(p^2)$ chiral Lagrangian is actually explained by 
including effects of $O(p^4)$ terms.

\section{A summary and discussions}
\label{sec:summary}

We studied the phase transition to the pion condensation phase for 
finite isospin chemical potential
using the holographic QCD model given in Refs.~\cite{EKSS,RP1}.
We introduced the isospin chemical potential $\mu_I$ 
as a UV-boundary value of the time component of 
the gauge field of SU(2)$_{\rm V}$ symmetry as
$V^3_0(z)|_\epsilon=\mu_I $.
We assumed non-existence of vector meson condensates since we are interested
in studying the small $\mu_I$ region.
Furthermore, we assumed that the neutral pion does not condense.
We solved the coupled equations of motion for the $\pi$-condensate and
``$\sigma$''-condensate together $V^3_0$ to determine $\mu_I$ as an eigenvalue.

Our result shows that the phase transition is of the second 
order and the critical chemical potential is predicted to be equal to 
the pion mass.
This is consistent with the result obtained by the chiral Lagrangian approach
in Ref.~\cite{SS}, but contrary to the result in Ref.~\cite{Albrecht:2010eg}. 
Furthermore,
our result on 
the relation between isospin number density and isospin chemical potential 
for small $\mu_I$ 
agrees with the one obtained by O$(p^2)$ chiral Lagrangian~\cite{SS}.
For large $\mu_I$ ($> 500$\,MeV), 
there is a difference between our predictions
and the one from O$(p^2)$ chiral Lagrangian, which is shown to be  understood as
the $O(p^4)$  contributions.

We also studied the $\mu_I$ dependence of the $\pi$-condensate and ``$\sigma$''-condensate.
Our result shows that, at the phase transition point, 
the $\pi$-condensate increases from zero as $\langle\pi^a\rangle \propto \left(\mu_I-\mu_I^c\right)^\nu$ with $\nu = 1/2$ consistently with the mean-field type of 
the phase transition.
Furthermore, we find that
the ``$\sigma$''-condensate decreases rapidly after the phase transition where the $\pi$-condensate grows rapidly, while 
the value of the "chiral condensate" defined by
$
\tilde{\langle \sigma\rangle} = \sqrt{ \langle \sigma \rangle^2 + \langle \pi^a \rangle^2 }
$
is constant for $\mu_I \lesssim 300$\,MeV,
and that it grows rapidly in the large $\mu_I$ region, which is contrary to the result by the strong coupling lattice shown in Ref.~\cite{Nishida:2004}.
This indicates that the chiral symmetry restoration at finite baryon density and/or finite temperature will be delayed when non-zero isospin chemical potential is turned on.  
\footnote{
Our result of the enhancement of the chiral symmetry breaking indicates that the critical point for the chiral phase transition may be shifted to higher chemical potential due to the existence of the isospin chemical potential.  This is consistent with the result of Ref.~\cite{Abuki:2013vwa}
.
}

In the present analysis, we did not include the effect of CS term.
When the CS term is included, 
an additional contribution from the $\omega$-type gauge field should be included.
However, as far as the O$(3)$ spatial rotation is kept unbroken, the result given in the present
analysis will not be changed.

It is interesting to study the $\rho$-meson condensation by extending the present analysis.
In such a case, the $\omega$-type gauge field will give a contribution through the CS term.
It is also interesting to include the explicit degrees of nucleons, by which we can study the 
phase structure including the baryon number chemical potential as well as the isospin
chemical potential.
This will be done by using the ``holographic mean field'' approach
done in Refs.~\cite{HaradaNakamuraTakemoto,HeHarada}.
We leave these analyses in future publications.

\section*{Acknowledgements}

We would like to thank Shin Nakamura for useful discussions and comments.
This work 
was supported in part by Grant-in-Aid for Scientific Research
on Innovative Areas (No. 2104) ``Quest on New
Hadrons with Variety of Flavors'' from MEXT,
and by the JSPS Grant-in-Aid for Scientific Research
(S) No. 22224003, (c) No. 24540266.

\end{document}